# Dielectric Anomalies in a New Manganocuprate, $Gd_3Ba_2Mn_2Cu_2O_{12}$

**Sudhindra Rayaprol[1,*], Anjana Dogra[2], S. D. Kaushik[1], Kiran Singh[3], S. Bhattacharya[4], G. Anjum[5], Y. Kumar[6], N. K. Gaur[3], and Ravi Kumar[6]**

[1]UGC-DAE Consortium for Scientific Research, Mumbai – 400085

[2]National Physical Laboratory, New Delhi – 110012

[3]Department of Physics, Barkatullah University, Bhopal – 462026

[4]TPPED, Bhabha Atomic Research Centre, Trombay, Mumbai – 400085

[5]Department of Physics, Aligarh Muslim University, Aligarh

[6]Inter University Accelerator Centre, New Delhi – 110067

## *Abstract*

Dielectric response has been studied for a new manganocuprate, $Gd_3Ba_2Mn_2Cu_2O_{12}$ (Gd3222) as a function of temperature (100 – 300 K) and frequency (75 kHz to 1 MHz). The dielectric constant (ε) exhibits a two step increase (two peaks) in $\varepsilon(T)$ with increasing temperature from 100 to 300 K. The first peak is seen around 150 K and the second one around 210 K (both for 75 kHz). Increasing frequency shifts both the peaks to higher temperature side. The behavior of dielectric constant (ε) and dielectric loss (tanδ) matches with glassy behavior observed in many dipolar molecules. 3D Variable range hopping type conduction was observed from the analysis of the resistivity data. The results obtained are discussed in light of similar dielectric behavior observed in charge-ordered manganites.

## 1. Introduction

The research on cuprate and manganite compounds has been on centre stage for last few decades. The novel physical properties of layered perovskite cuprate and manganite compounds have led to volumes of work on basic sciences as well as on the technological front [1, 2]. Several years since their discoveries, the research on these two systems continues unhindered since both the systems offer great freedom in exploring crystal chemistry as well as modifying physical properties through chemical control (by substitutions, doping etc). This on one hand enriched the solid-state chemistry of layered oxide compounds, and on the other hand gave rise to new compounds with interesting physical properties. The rare earth manganites, $Ln\mathrm{MnO}_3$ exhibits colossal magnetoresistance when doped with electron or holes (where *Ln* = trivalent lanthanide ion). Several electron doped $LnMnO_3$ (*n*-type) perovskites e.g., $Ca_{1-x}Ln_xMnO_3$ are very good thermoelectric materials also. In recent times, many studies are devoted in studying multiferroic properties of $LnMnO_3$ type compounds [3]. Many charge ordered manganite systems, typically of the type $Ln_{0.5}A_{0.5}MnO_3$ (*A* = divalent alkali metal ions) have exhibited dielectric (multiferroic) properties [3, 4].

Recently it was found that the cubic-perovskite related compound, $CaCu_3Ti_4O_{12}$ (CCTO) exhibits *colossal dielectric constant* (CDC) [5]. The term CDC is used since CCTO exhibits almost temperature independent dielectric constant (ε) in the temperature range 100 – 600 K, which drops almost by order of *3* below 100 K. The observation of dielectric properties in cuprates containing transition metal ions like CCTO opens up a new avenue in the search for novel materials exhibiting multiferroic properties.

Hervieu *et al* [6] synthesized a new manganocuprate compound, $Eu_3Ba_2Mn_2Cu_2O_{12}$ which is a made up of the blocks of $Eu_2MnO_4$ ($K_2NiF_4$ type) separated by layered perovskite $EuBa_2Cu_2MnO_8$ (*R*123 type). The structural determination of $Eu_3Ba_2Mn_2Cu_2O_{12}$ has been done in detail by Hervieu *et al* and Field *et al* [6, 7] using high resolution electron microscopy (HREM) and powder neutron diffraction (PND) respectively. From the analysis of the resistivity data of Eu3222, activation energy ($E_a$) of 0.17 eV was observed in the temperature range of 100 – 300 K indicating the localization of the charge carriers [6]. The insulating behavior and localization of charge carries rule out the mixed valence state for Cu, therefore the most likely possibility of oxidation states suggests that Mn is in two valence states $Mn^{3+}$ and $Mn^{4+}$. From the HREM studies

Hervieu *et al* concludes that among various possibilities of oxidation states for Mn and Cu, the most likely (or closest) oxidation states for the ions would be given as:

$$Eu_3^{3+}Ba_2^{2+}Mn^{3+}Mn^{4+}Cu_2^{2+}O_{12}^{2-} \quad \text{---- (1)}$$

The presence of $Mn^{3+}/^{4+}$ states presents a scenario similar to the ones observed usually in manganites. For a typical half-doped manganite exhibiting charge-ordering, the oxidation states can be exhibited as

$$(Ln_{0.5}^{3+}A_{0.5}^{2+})(Mn_{0.5}^{3+}Mn_{0.5}^{4+})O_3^{2-} \quad \text{--- (2)}$$

The equal distribution of $Mn^{3+}$ and $Mn^{4+}$ valence states is believed to give rise to the charge-ordering in the half-doped manganites [2, 9].

Therefore, on summarizing these observations, i.e., (i) CDC was observed in a copper-titanate system (CCTO), (ii) equal distribution of $Mn^{3+/4+}$ in Eu3222, similar to the charge-ordered manganites; the studies on manganocuprates (*Ln*3222) promises exciting novel physical properties.

Our recent study on Gd3222 establishes that it is an isostructural compound of $Eu_3Ba_2Mn_2Cu_2O_{12}$ [8]. The low temperature studies on Gd3222 exhibits anomalous magnetic behavior and a spin-glass like state at temperatures below 10 K. Gd3222 also exhibits a magnetocaloric effect (MCE) of about 27 J/mol K at 10 K under a field change of 10 Tesla [8]. Encouraged by the initial results on the manganocuprates we present here the resistivity behavior of Gd3222 measured as a function of temperature. We have also studied the dielectric response of Gd3222 in a temperature range of 100 – 300 K at various frequencies between 75 kHz to 1 MHz. The results are discussed to exhibit the multi-functional properties of Gd3222 compound and to study the inter-relationship among various phenomenons.

## 2. Experimental Details

The polycrystalline samples were prepared by solid-state reaction method using high purity (minimum stated purity > 99.9%) starting compounds of $Gd_2O_3$, $BaCO_3$, $MnO_2$ and CuO. X-ray diffraction (XRD) patterns were recorded for the powdered samples to check the phase purity. The title compound crystallizes in a tetragonal unit cell, space group *I4/mmm*. Details of sample synthesis and XRD are given by Rayaprol *et al* [8]. Resistivity for the Gd3222 compound was measured using the four probe geometry on a closed cycle refrigerator (CCR) based cryostat. The dielectric measurements of both the systems were made using HP LCR meter (model No. *Agilent* 4285A) in the temperature range of 100 K–300 K at certain selected frequencies.

## 3. Results and discussion

### 3.1 *Resistivity measurements*

In Fig. 1 the resistivity ($\rho$) as a function of temperature for Gd3222 sample in the temperature ranges 100 – 300 K is presented. The room temperature resistivity value for Gd3222 was 65Ωcm, which increases with decreasing temperature, exhibiting semiconductor like behavior. In the study of transport properties of layered manganite systems, the semiconducting behavior of resistivity is explained by several mechanisms such as small polaron hopping, Mott's 2D or 3D variable range hopping. Gupta *et al* [10] has recently showed that from the analysis of the variation of activation energy ($E_a$) with respect to temperature, the mechanism responsible for conduction in layered manganites can be proposed. According to Sarathy *et al* [11] the temperature dependence of $E_a$ can be expressed as:

$$E_a = -\frac{k_B T^2}{\rho}\left(\frac{d\rho}{dT}\right) \qquad \text{--- (3)}$$

Where, $E_a$ = activation energy, $k_B$ = Boltzmann's constant, $\rho$ = resistivity.

A general expression for elucidating the mechanism of conduction can be given as:

$$\rho = \rho_0 \exp\left(\frac{B}{T}\right)^{\gamma} \qquad \text{--- (4)}$$

Where B = constant, $E_a/k_B$

$\gamma$ = 1 exhibits nearest neighbor hopping,

= ¼ exhibits Mott's 3D variable range hopping,

= ½ exhibits Mott's 2D variable range hopping, and

= ⅓ exhibits Efros-Shklovskii's variable range hopping

Therefore combining equations (3) and (4), the expression for $E_a(T)$ can be given as

$$E_a = \varepsilon T^{\delta} \text{; where } \varepsilon = \gamma B^{\gamma} k_B \text{ and } \delta = 1 - \gamma = 3/4 \qquad \text{--- (5)}$$

For 3D Mott variable range hopping the expression is given as

$$E_a = \varepsilon T^{3/4} \qquad \text{--- (6)}$$

The relation between $\varepsilon$ and Mott's parameter $T_0$ is given as

$$\varepsilon = \frac{1}{4}(T_0)^{1/4} k_B \qquad \text{--- (7)}$$

In figure 2, the activation energy calculated using equation (3), is plotted against the temperature of measurement. The solid line passing through the calculated $E_a$ data is the best fitted from the equation (5). The value of $\delta$ obtained from the fit is 0.72, and very close to the expected value of 3/4 (= 0.75) for Mott's 3D variable range hopping mechanism (see equations (6) and (7) also). The inset in Fig. 1 shows the plot of $\rho$ *vs.* $1/T^{1/4}$, i.e., the three dimensional variable range hopping (VRH) model. The solid line passing through the data points is the linear fit of the data. The linearity of the plot shows that the conduction mechanism in Gd3222 is governed by hopping of charges in 3-dimensions.

The Mott's 3D variable range hopping is expressed as:

$$\rho = \rho_0 \exp\left(\frac{T_0}{T}\right)^{1/4} \qquad \text{--- (8)}$$

where $T_0$ = the Mott's parameter.

Sarathy *et al* has shown that using the values of $T_0$, we can calculate the density of states, $N(E_F)$ using the relation,

$$T_0 = \frac{16\alpha^3}{k_B N(E_F)} \qquad \text{--- (9)}$$

Where $\alpha$ (numerical coefficient) is taken as 2.5/nm. Using the equation (9), the calculated density of state for Gd3222 is of the order of $10^{20}\ eV^{-1}\ cm^{-3}$. This value is close to the values expected for charged ordered manganites and bi-layered manganites exhibiting large magnetoresistance [10, 11].

3.2 *Dielectric measurements*

The dielectric response of Gd3222 as a function of temperature is shown in Fig. 3. The figure exhibits two plateaus (peaks in first derivative) around 150 and 210 K for 75 MHz. With increasing frequency, the peak magnitude of the peak decreases marginally, however the plateau (or say peak for clarity) shifts to higher temperature on increasing the frequency. Similarly *tanδ* (Fig. 4) also exhibits peaks corresponding to *ε(T)* and they also exhibit frequency dependence. The frequency dependence of *ε(T)* exhibits dielectric relaxation, and this can be explained by Vogel-Fulcher (VF) law, given as:

$$\nu = \nu_0 \exp\left(\frac{-E_a}{k_B(T_p - T_a)}\right) \qquad \text{--- (10)}$$

Where, $\nu_0$ is the pre-exponential factor or also known as the characteristic frequency ($\nu_0 = 10^{12}$ Hz), $E_a/k_B$ represents the activation barrier and $T_a$ is known as the freezing temperature. $T_p$ is the temperature of maxima in $\varepsilon(T, \nu)$ plot. In Fig. 5, we have plotted the VF relation for both peaks of Gd3222. The straight line passing through the data lines is the linear fit of the data points, exhibiting that the dielectric relaxation in the manganocuprates are glassy type. However, it must be cautioned here that, to completely elucidate the relaxation behavior in Gd3222 more detailed dielectric results are required and also the contribution of intrinsic and extrinsic effects to the dielectric constants should also be evaluated.

## 4. Conclusions

A new manganocuprate have been studied for its dielectric behavior at various frequencies in the temperature range 100 – 300 K. The dielectric constant ($\varepsilon$) increases by two orders of magnitude in the case of Gd3222. The plot of $\varepsilon(T, \nu)$ exhibits relaxation peaks which are frequency dependent. The dielectric dispersion of the manganocuprates is similar to that of manganite systems exhibiting charge-ordering. The resistivity data for Gd3222 also establishes similarity in the conduction mechanism between the manganocuprates as well as charge-ordered manganites. Therefore, it would be interesting to extend the present studies on the lines of perovskite manganites to understand the underlying physics of this interesting compound.

**Figures Captions**

Figure 1 (colour online) Resistivity ($\rho$) as a function of temperature for $Gd_3Ba_2Mn_2Cu_2O_{12}$. The linear variation of ρ as a function of $T^{-1/4}$ is shown in the inset.

Figure 2 The temperature dependence of activation energy ($E_a$) for $Gd_3Ba_2Mn_2Cu_2O_{12}$, calculated from the resistivity data (shown in Fig. 1). The straight line passing through the data points ($175 < T(K) < 240$) is the best fit of the equation (5). *See text for details*.

Figure 3 (colour online) Dielectric constant ($\varepsilon$) as a function of temperature, measured at different frequencies for $Gd_3Ba_2Mn_2Cu_2O_{12}$.

Figure 4 (colour online) Dielectric loss ($tan\ \delta$) as a function of temperature, measured at different frequencies for $Gd_3Ba_2Mn_2Cu_2O_{12}$.

Figure 5 The Vogel-Fulcher plot for $Gd_3Ba_2Mn_2Cu_2O_{12}$. The straight lines passing through the data points are the linear fits according to equation (10).

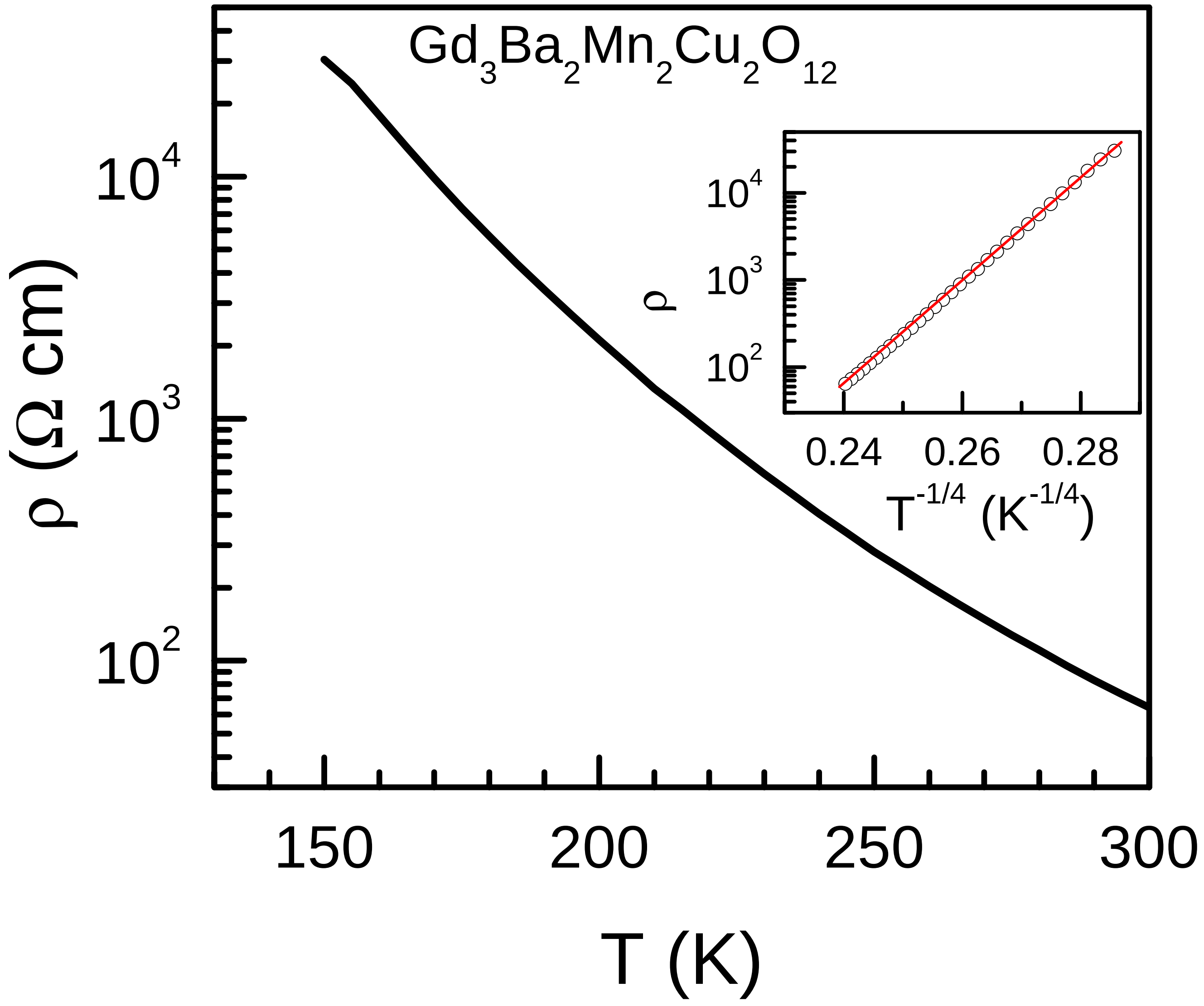

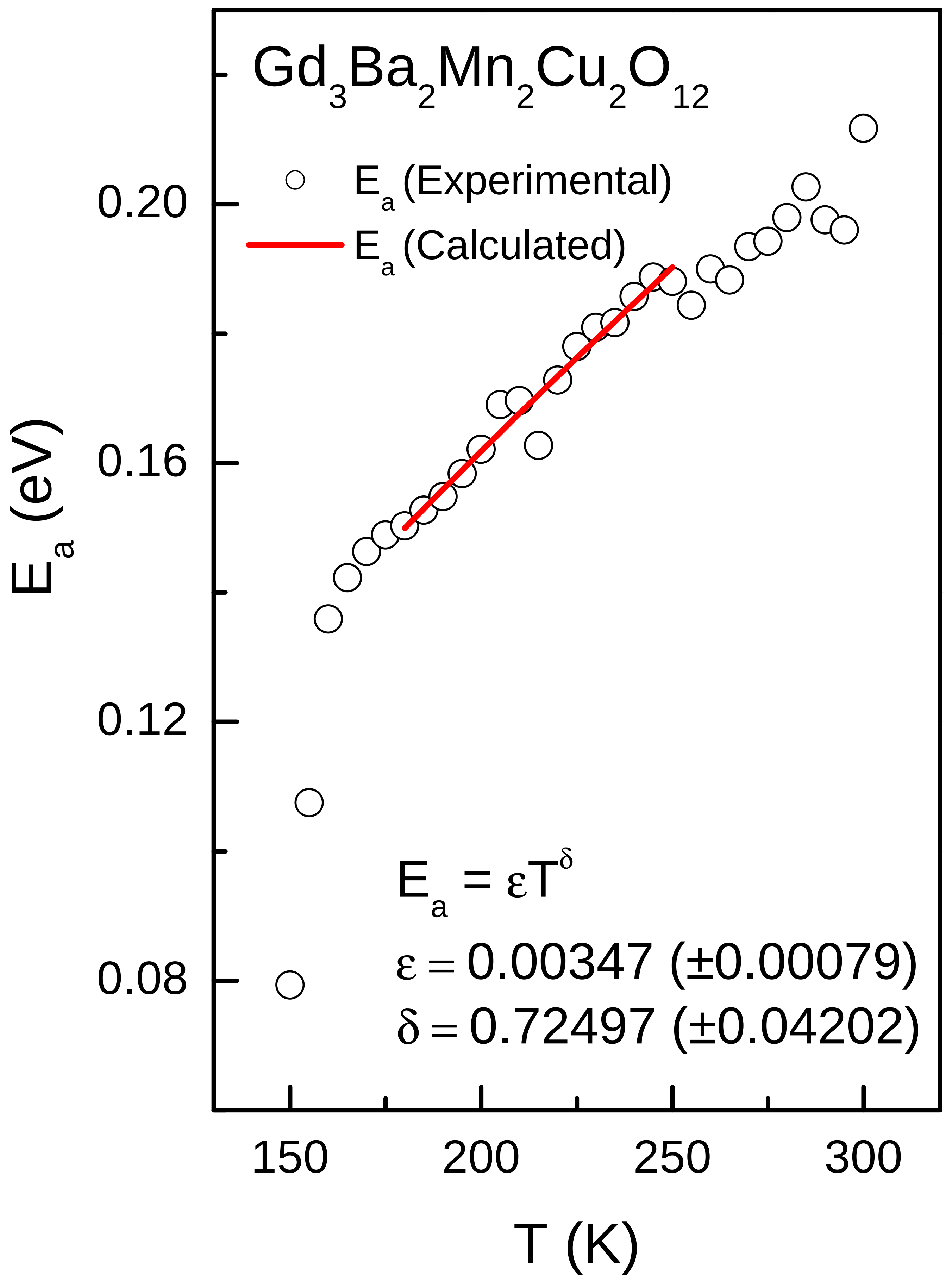

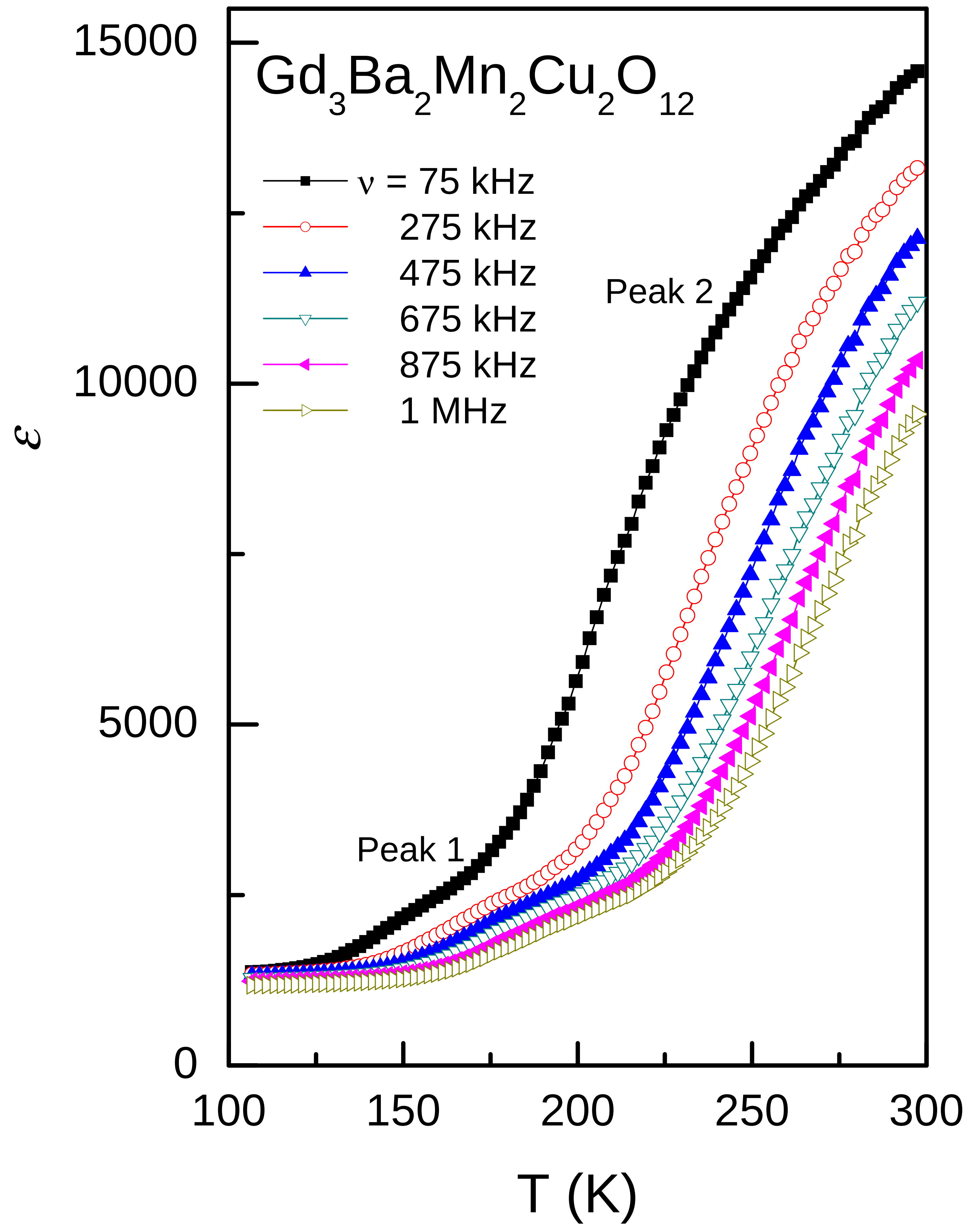

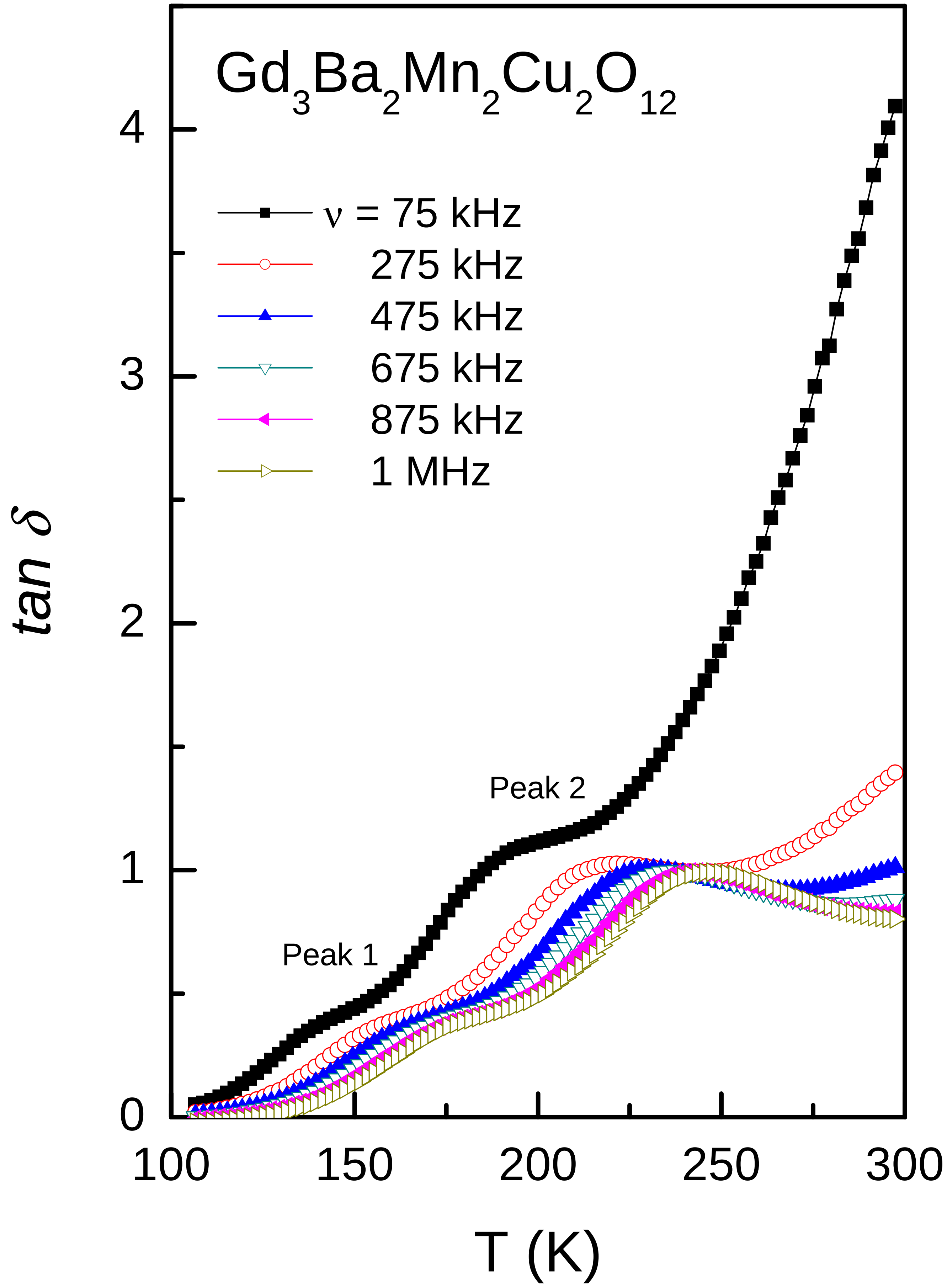

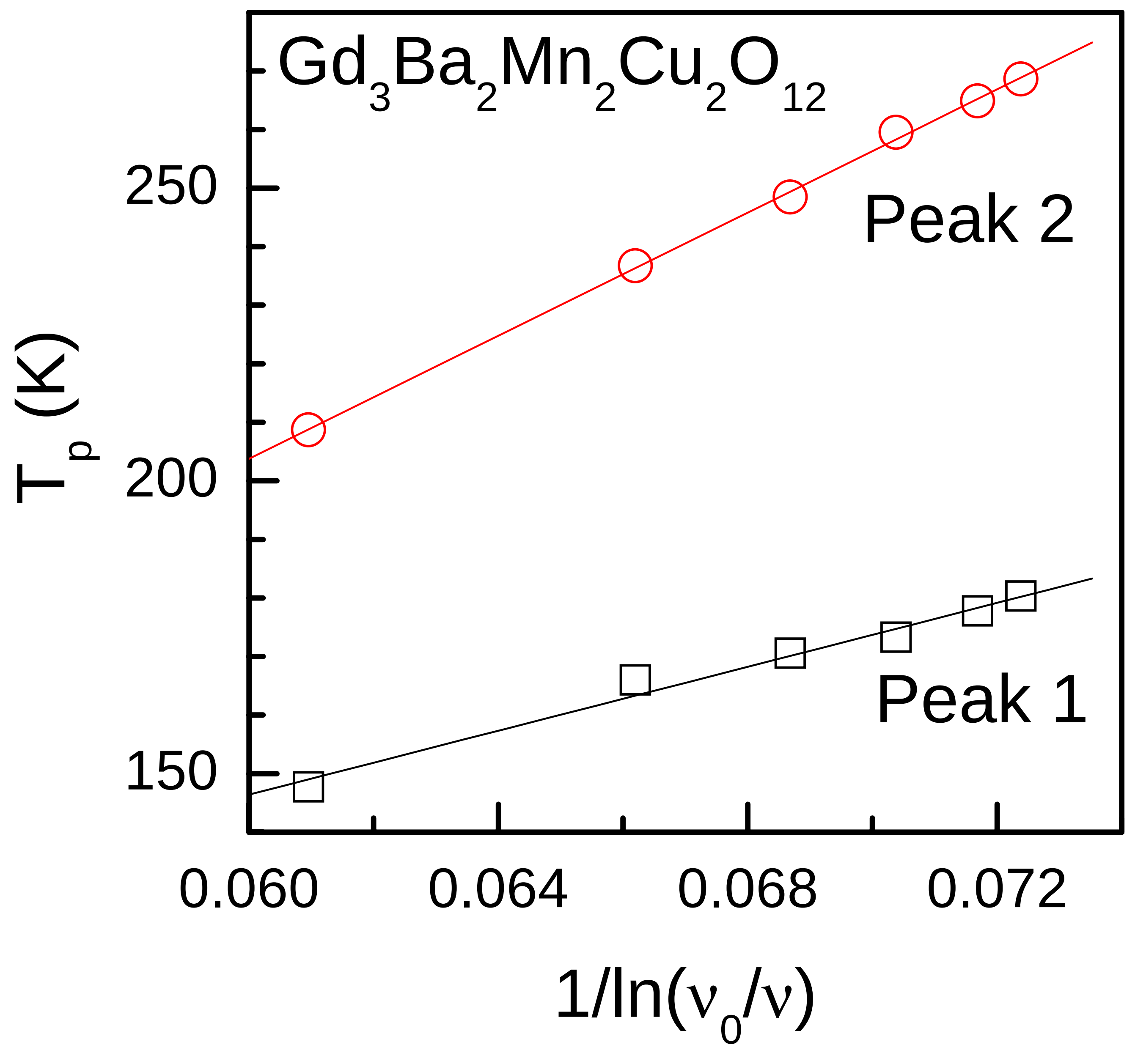